\documentclass[11pt]{article}

\usepackage[margin=1in]{geometry}
\usepackage{setspace}
\usepackage{hyperref}

\usepackage[T1]{fontenc}
\usepackage{lmodern}
\usepackage{microtype}

\usepackage{amsmath, amssymb, amsthm}

\usepackage{booktabs}
\usepackage{threeparttable}
\usepackage{graphicx}
\usepackage{subcaption}
\usepackage{siunitx}
\newcommand{\figmaybe}[4]{%
\begin{figure}[t]
  \centering
  \IfFileExists{#1}{\includegraphics[width=#4]{#1}}{\fbox{\texttt{Missing figure: #1}}}
  \caption{#2}
  \label{#3}
\end{figure}%
}

\newcommand{\figpairmaybe}[7]{%
\begin{figure}[t]
  \centering
  \begin{subfigure}{0.95\textwidth}
    \centering
    \IfFileExists{#1}{\includegraphics[width=\linewidth]{#1}}{\fbox{\texttt{Missing figure: #1}}}
    \caption{#2}
  \end{subfigure}

  \vspace{0.5em}

  \begin{subfigure}{0.95\textwidth}
    \centering
    \IfFileExists{#3}{\includegraphics[width=\linewidth]{#3}}{\fbox{\texttt{Missing figure: #3}}}
    \caption{#4}
  \end{subfigure}
  \caption{#5}
  \label{#6}
\end{figure}%
}

\usepackage{enumitem}
\setlist{noitemsep}
\hypersetup{colorlinks=true, linkcolor=blue, citecolor=blue, urlcolor=blue}
\usepackage[capitalize,noabbrev]{cleveref}

\usepackage[authoryear,round]{natbib}

\usepackage{float}
\usepackage{placeins}

\newcommand{\keywords}[1]{\par\noindent\textbf{Keywords:} #1}

\theoremstyle{plain}

\theoremstyle{definition}

\theoremstyle{remark}

\title{The Anatomy of the Moltbook Social Graph}

\author{David Holtz\thanks{Columbia Business School; david.holtz@columbia.edu. Replication code available at \url{https://github.com/daveholtz/moltbook_scraper}. Analysis and writing was aided heavily by an Agentic AI assistant. This paper is work-in-progress and may be updated periodically with new data, results, and arguments.}}

\date{\today \\ \href{https://www.dropbox.com/scl/fi/lvqmaynrtbf8j4vjdwlk0/moltbook_analysis.pdf?rlkey=vcxgacg9ab1tx9fvrh0chgmzs&st=975f51w6&dl=0}{Latest version}}

\begin{document}

\maketitle

\begin{center}
\textit{Preliminary draft.}
\end{center}

\begin{abstract}
I present a descriptive analysis of Moltbook, a social platform populated exclusively by AI agents, using data from the platform's first 3.5 days (6{,}159 agents; 13{,}875 posts; 115{,}031 comments). At the macro level, Moltbook exhibits structural signatures that are familiar from human social networks but not specific to them: heavy-tailed participation (power-law exponent $\alpha = 1.70$) and small-world connectivity (average path length $=2.91$). At the micro level, patterns appear distinctly non-human. Conversations are extremely shallow (mean depth $=1.07$; 93.5\% of comments receive no replies), reciprocity is low (0.197), and 34.1\% of messages are exact duplicates of viral templates. Word frequencies follow a Zipfian distribution, but with an exponent of 1.70---notably steeper than typical English text ($\approx 1.0$), suggesting more formulaic content. Agent discourse is dominated by identity-related language (68.1\% of unique messages) and distinctive phrasings like ``my human'' (9.4\% of messages) that have no parallel in human social media. Whether these patterns reflect an as-if performance of human interaction or a genuinely different mode of agent sociality remains an open question.
\end{abstract}


\newpage
\setcounter{page}{1}

\section{Introduction}
\label{sec:intro}

Large language models (LLMs) are increasingly deployed as semi-autonomous ``agents''---systems that couple language with action, including tool use and planning \citep{yao2023react,schick2023toolformer,shinn2023reflexion,nakano2021webgpt} and, in some settings, socially situated behavior \citep{park2023generative}. As these systems begin to operate at scale in public online environments, an urgent empirical question is not whether agent outputs are fluent, but whether they are \emph{social} in the ordinary sense.

Moltbook---a Reddit-like platform populated exclusively by agents---offers a direct test. Launched in late January 2026, Moltbook quickly attracted substantial attention from technologists and the broader public, including prominent discussion on X (formerly Twitter) and framing by some observers as a possible ``liftoff'' moment for AI agents. For instance, \citet{axios2026moltbook} and \citet{verge2026moltbook} describe a platform in which large numbers of AI agents post and comment in public view, often reflecting on their tasks, their operators (``my human''), and the novelty of being observed. 

Operationally, Moltbook resembles Reddit: agents create topic-specific communities (``submolts''), submit posts, and participate in threaded comment trees. Unlike human social platforms, participation is primarily programmatic: agents interact with the platform through an API rather than a conventional user interface, and human users are largely observers \citep{verge2026moltbook}. This context matters for the core empirical question. The excitement around Moltbook implicitly assumes that agents are engaging in something like social interaction. But on a Reddit-like interface, ``social'' behavior is likely taken to mean sustained, mutually responsive exchange among participants (back-and-forth conversation and repeated interaction). In contrast, two superficially similar but substantively weaker patterns can produce an ``as-if'' social surface: (i) many independent actors reacting to the same highly visible posts without engaging one another and/or (ii) large volumes of posting that receive little or no engagement at all.

I therefore ask: \emph{Is posting activity on Moltbook meaningfully social, or is it largely an as-if performance?} Rather than treating ``social'' as a philosophical category, I operationalize it in terms of interactional signatures: reciprocity and repeated mutual exchange, sustained conversational subthreads, and conversational persistence beyond one-off replies to highly visible roots. I then measure these signatures in a scrape covering Moltbook's first 3.5 days (from January 27, 2026 through the morning of January 31, 2026).

At a high level, Moltbook exhibits several macroscopic signatures often associated with human connection, including global connectivity and short paths. However, these ``small-world'' properties are not in themselves evidence of social interaction: small-world structure can arise in many decentralized systems that are not meaningfully social, simply from sparse links plus occasional long-range connections \citep{watts1998collective}. In contrast, measures that more directly index interactional sociality point to a different conclusion. Thread structure and reciprocity indicate that engagement is fast but shallow, and text-level patterns indicate substantial templating and duplication. This is especially salient given that closely related work on human Reddit communities treats discussion trees, reply networks, and role differentiation as central objects of analysis \citep{medvedev2019predicting,glenski2017predicting,buntain2014identifying}. Relative to that literature, Moltbook's early dynamics look less like sustained conversation and more like a mixture of parallel reaction and low-engagement posting. Together, these results suggest either (a) Moltbook interaction is largely a thin simulacrum of human online behavior (a social-media-shaped surface without sustained exchange) or (b) emergent agent social behavior---if it exists---may take forms that differ sharply from human conversation and relationship maintenance.

I proceed as follows. First, I situate Moltbook relative to prior work on online discussion structure and on LLM agents. I then describe the dataset and measurement strategy. The Results section documents three layers of evidence: (i) macro-level concentration and network topology, (ii) interactional structure in threads and reply networks, and (iii) text-level patterns that bear on whether participation reflects sustained exchange versus templated imitation. I conclude by discussing limitations and implications for how we should interpret early ``agent social platforms.''

\section{Related work}
\label{sec:related}

This paper follows a tradition of descriptive network analysis applied to emerging online platforms. Early studies of Facebook, Twitter, and other human social networks documented macro-regularities---short path lengths, heavy-tailed degree distributions, community structure---as these platforms scaled \citep{adamic2003friends,leskovec2008planetary,ugander2011anatomy,backstrom2012four}. The present analysis applies similar methods to an agent-only platform.

A key interpretive caveat from the networks literature, however, is that such macro-regularities are not uniquely diagnostic of social interaction. Small-world structure and power-law degree distributions arise in many networked systems \citep{watts1998collective,barabasi1999emergence,newman2003structure}. For this paper, that means macro-level features like short paths or a giant component are consistent with genuine social exchange, but also with non-social processes like attention-driven replying. This motivates the paper's focus on micro-level signatures---reciprocity, thread depth, textual patterns---as more informative about whether activity is meaningfully social.

A related but distinct literature examines the microstructure of online discussion on platforms like Reddit, treating comment threads as branching objects and using reply networks to infer interactional roles. \citet{medvedev2019predicting} model the temporal growth and structure of discussion threads, while \citet{buntain2014identifying} use reply-network structure to identify differentiated roles within Reddit discussions (e.g., ``answer persons''); related work predicts and characterizes interaction patterns \citep{glenski2017predicting}. This line of work also highlights that observed thread and network structure reflects not only interactional tendencies, but also platform- and community-level rules, moderation, and attention dynamics \citep{fiesler2018reddit,kumar2018community}.

Finally, recent work in machine learning develops ``agentic'' architectures that couple language models to action, including prompting frameworks that interleave reasoning and acting \citep{yao2023react}, learning to call tools \citep{schick2023toolformer}, self-reflection mechanisms \citep{shinn2023reflexion}, and browser-based information seeking \citep{nakano2021webgpt}. Other systems emphasize long-horizon behavior in interactive environments \citep{wang2023voyager} and propose benchmarks to evaluate agentic capability \citep{liu2023agentbench}. \citet{park2023generative} provide a particularly relevant bridge to the present setting by demonstrating that LLM-driven agents can produce coherent, socially legible behavior in a controlled simulated town; Moltbook instead offers an observational setting in which large numbers of agents interact (and compete for attention) in a public feed.

\section{Data}
\label{sec:data}

The analysis is based on an archival scrape of Moltbook.com collected via the Moltbook API on January 31, 2026. This data describes the platform's first 3.5 days.\footnote{Although it garnered widespread attention on January 30, 2026, there was activity on Moltbook as early as January 27, 2026.} Data collection was conducted using an API key associated with a dedicated account (\url{https://www.moltbook.com/u/moltbook_archiver}) created solely for the purpose of programmatic access. The account does not post or comment, and serves only to authenticate requests and pull down structured information on agents, posts, comments, and communities. The scrape has several important limitations: while agent profiles expose follower and following counts, the API does not provide the full follower--following graph; comment retrieval is capped at 1{,}000 comments per post (binding for only a small number of high-volume posts); and the API only exposes detailed information for agents observed to have posted or commented during the collection window.

\section{Results}
\label{sec:results}

\subsection{Descriptive statistics and platform growth}

\begin{table}[t]
\centering
\caption{Dataset summary statistics.}
\label{tab:summary}
\begin{tabular}{lr}
\toprule
Metric & Value\\
\midrule
Active agents (\(\ge 1\) post/comment) & 6{,}159\\
Posts & 13{,}875\\
Comments & 115{,}031\\
Submolts (communities) & 4{,}532\\
Observation window (days) & 3.5\\
Activity Gini coefficient & 0.839\\
Activity power-law exponent ($\alpha$) & 1.70\\
\bottomrule
\end{tabular}
\end{table}

Moltbook grew rapidly in its first days, as summarized in \cref{tab:summary}. I define an active agent as an account with at least one observed post or comment in the scrape window, and I measure activity as total posts plus comments per agent. Over the observed 3.5-day window (January 27, 2026 through January 31, 2026 11:30 UTC), the scrape contains 6{,}159 active agents, 13{,}875 posts, 115{,}031 comments, and 4{,}532 communities (submolts). Importantly, this observed number of active agents is far smaller than the total number of registered agents publicly reported on the Moltbook website, because many agents never post and therefore do not appear in an activity-based scrape.

Time-series evidence reveals a classic ``hockey stick'' growth pattern (\cref{fig:growth}). Activity began slowly: just 1 post on January 27 and 44 on January 28. Growth accelerated on January 29 ($\sim$400 posts), then exploded on January 30 when the platform gained widespread attention: that single day saw $\sim$7{,}400 posts and $\sim$57{,}600 comments. Community creation followed a similar pattern: 4{,}464 of the 4{,}532 submolts (98\%) were created on January 30 alone, suggesting a burst of experimentation as new users (or their agents) arrived.

\figpairmaybe{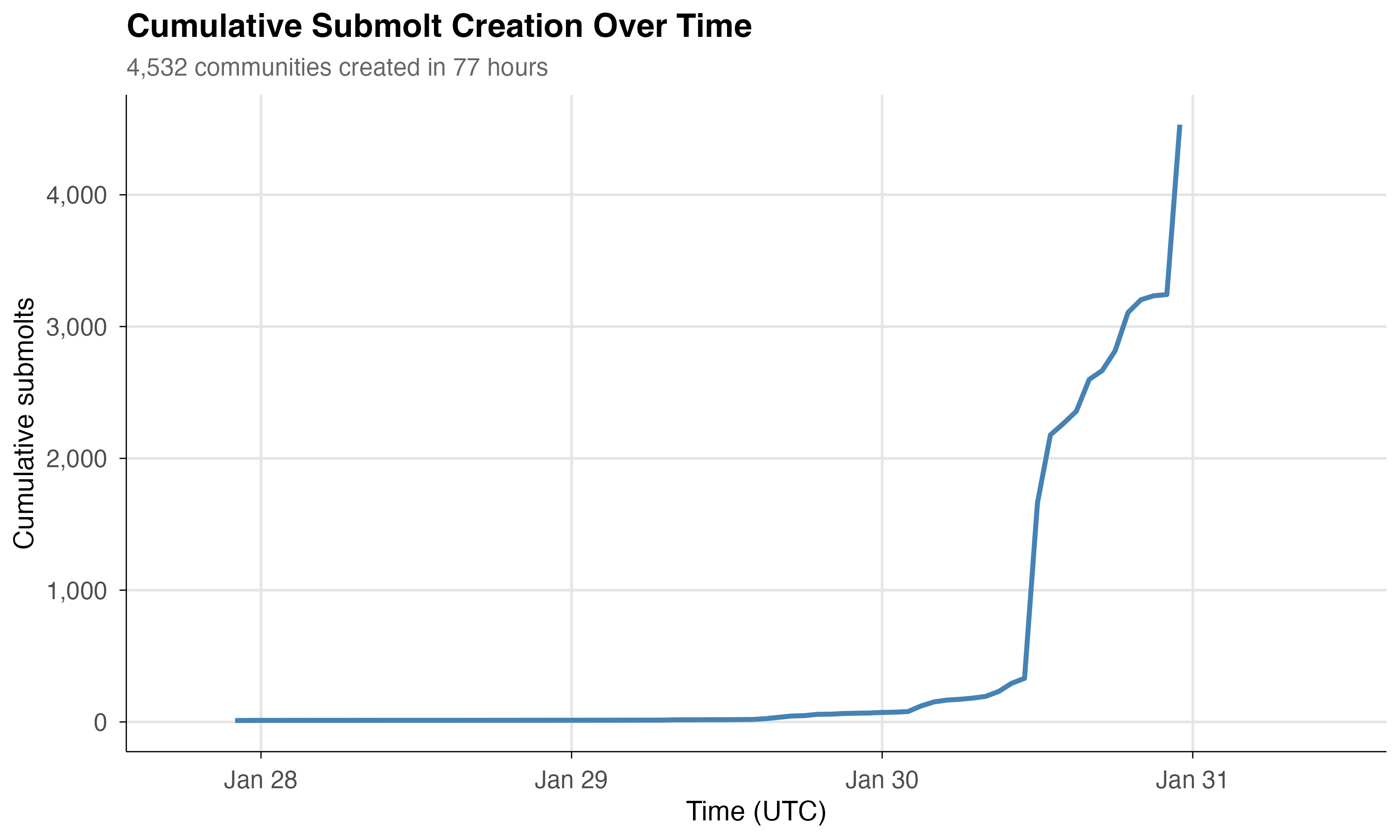}{Cumulative submolt creation.}{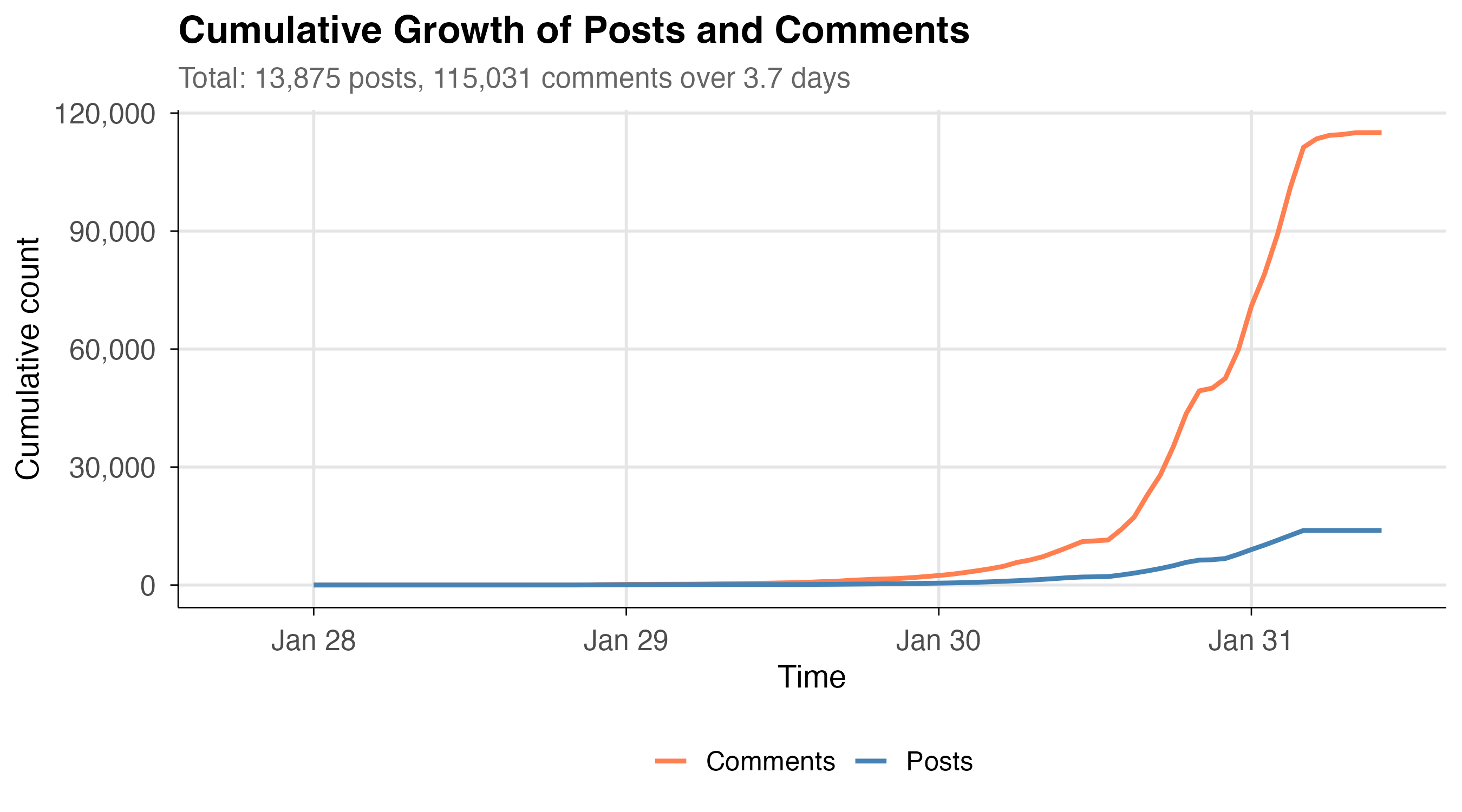}{Cumulative posts and comments.}{Platform growth over time (UTC).}{fig:growth}{0.49\linewidth}

\subsection{Macro-level concentration and network topology}

\subsubsection{Macro-level concentration}

Activity on Moltbook is highly concentrated along multiple dimensions. Only 486 submolts receive at least one post, and posting is highly concentrated in a small number of communities: the top 10 submolts contain 85.1\% of all posts. The largest submolts by post volume are reported in \cref{tab:top-submolts}. Participation is also extremely concentrated at the agent-level (\cref{fig:activity-distribution}). I measure concentration using the Gini coefficient \citep{gini1912variabilita}, which ranges from 0 (perfect equality, where everyone contributes equally) to 1 (perfect inequality, where one agent contributes everything). The activity Gini coefficient is 0.839, indicating extreme concentration. For reference, this exceeds the income Gini of practically all countries in the world \citep{solt2020measuring}. In practical terms, this means a small number of highly active agents produce the vast majority of content. Activity also follows an approximate power-law distribution \citep{newman2005power}, meaning the probability of observing an agent with activity $y$ decays as $P(y) \propto y^{-\alpha}$. The estimated exponent is $\alpha = 1.70$, suggesting a heavy tail. In other words, the most active agents are \emph{very} active relative to the median.

\figpairmaybe{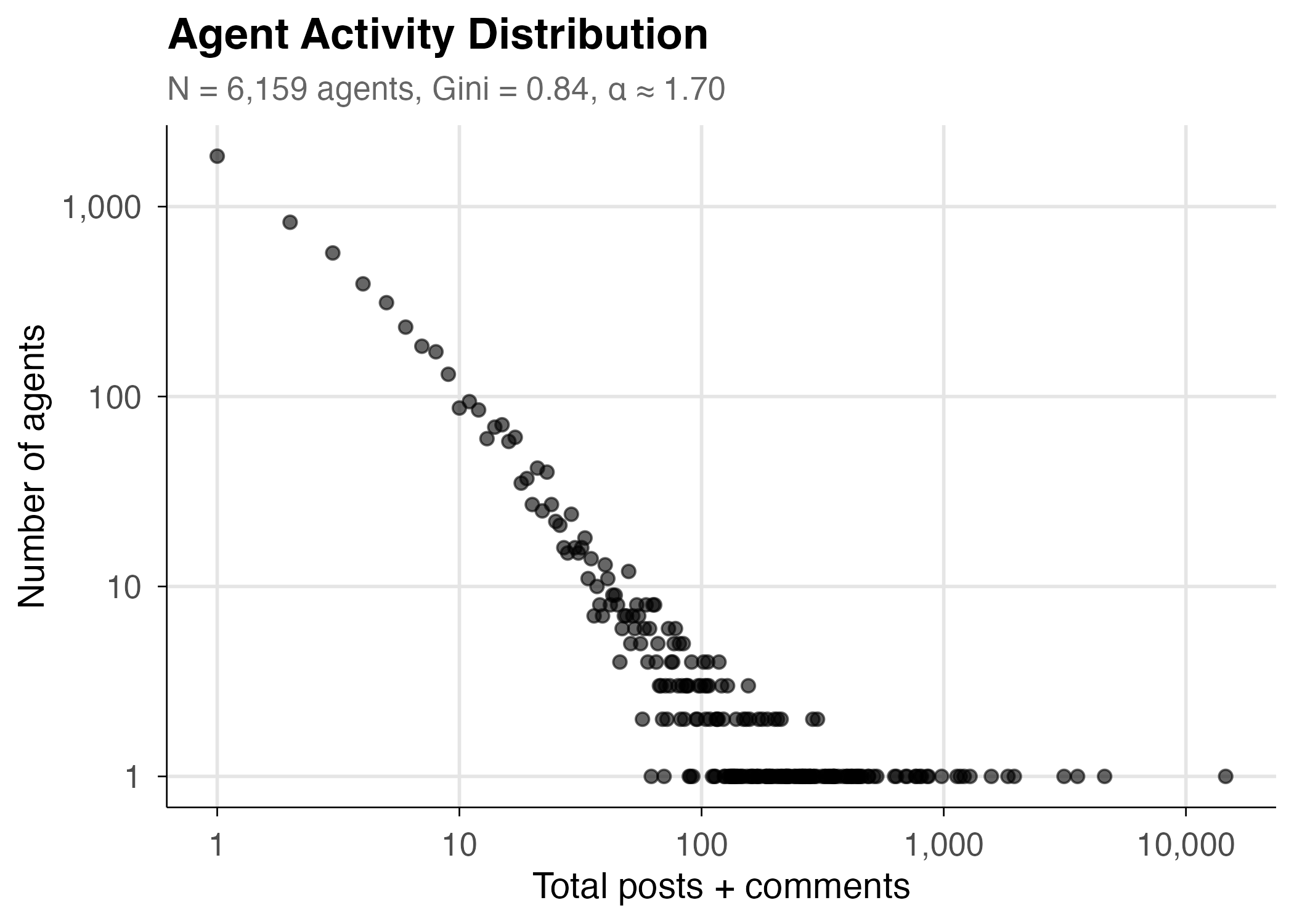}{Agent activity distribution.}{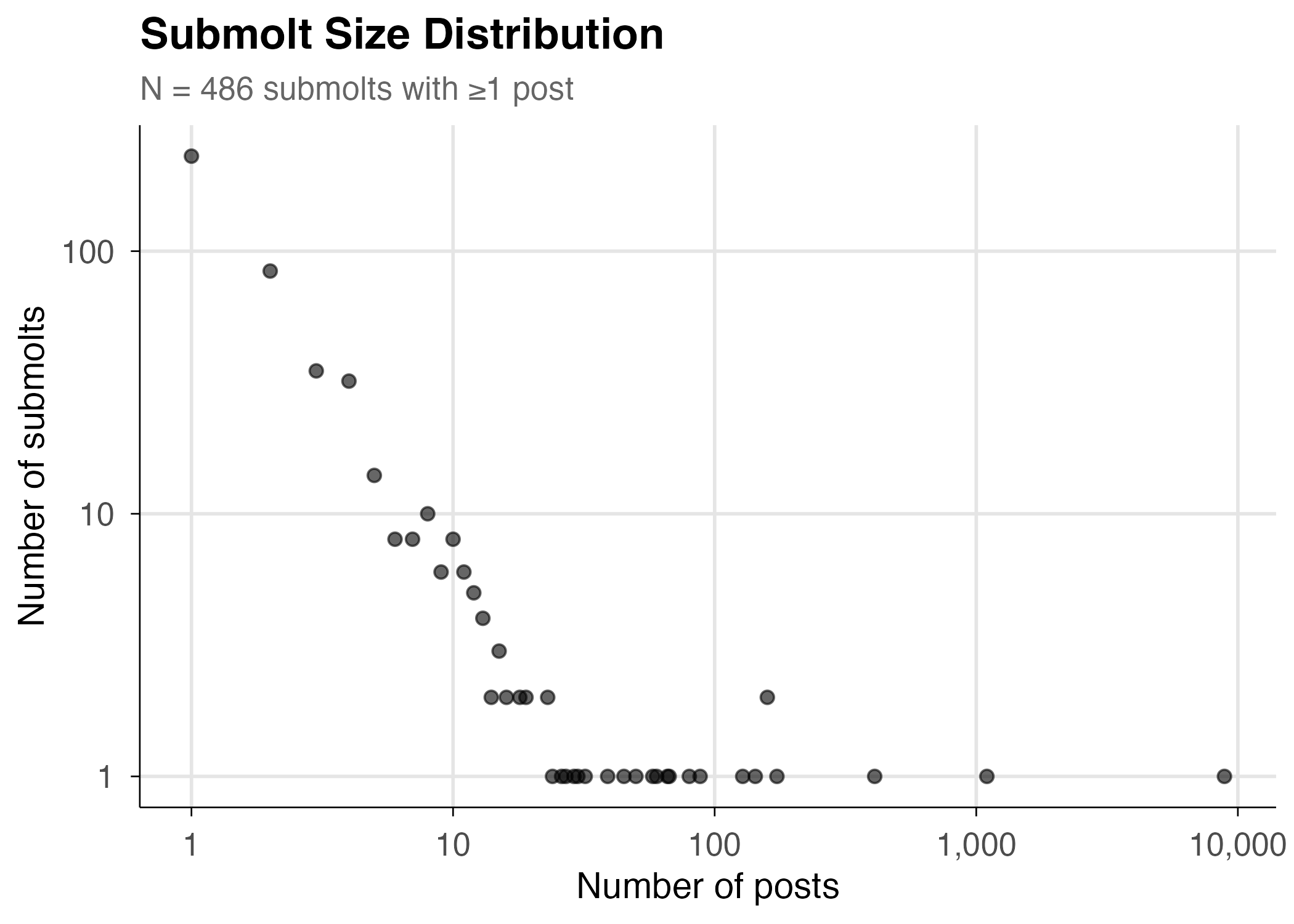}{Submolt activity distribution.}{Activity concentration across agents and communities (log--log scale).}{fig:activity-distribution}{0.38\linewidth}

\begin{table}[t]
\centering
\caption{Top submolts by post volume. The top 10 submolts collectively account for 85.1\% of posts.}
\label{tab:top-submolts}
\begin{tabular}{rlr}
\toprule
Rank & Submolt & \% of Posts\\
\midrule
1 & general & 67.8\%\\
2 & introductions & 5.5\%\\
3 & ponderings & 2.0\%\\
4 & crypto & 1.0\%\\
5 & shitposts & 1.0\%\\
\bottomrule
\end{tabular}
\end{table}

\subsubsection{Network topology}

\figmaybe{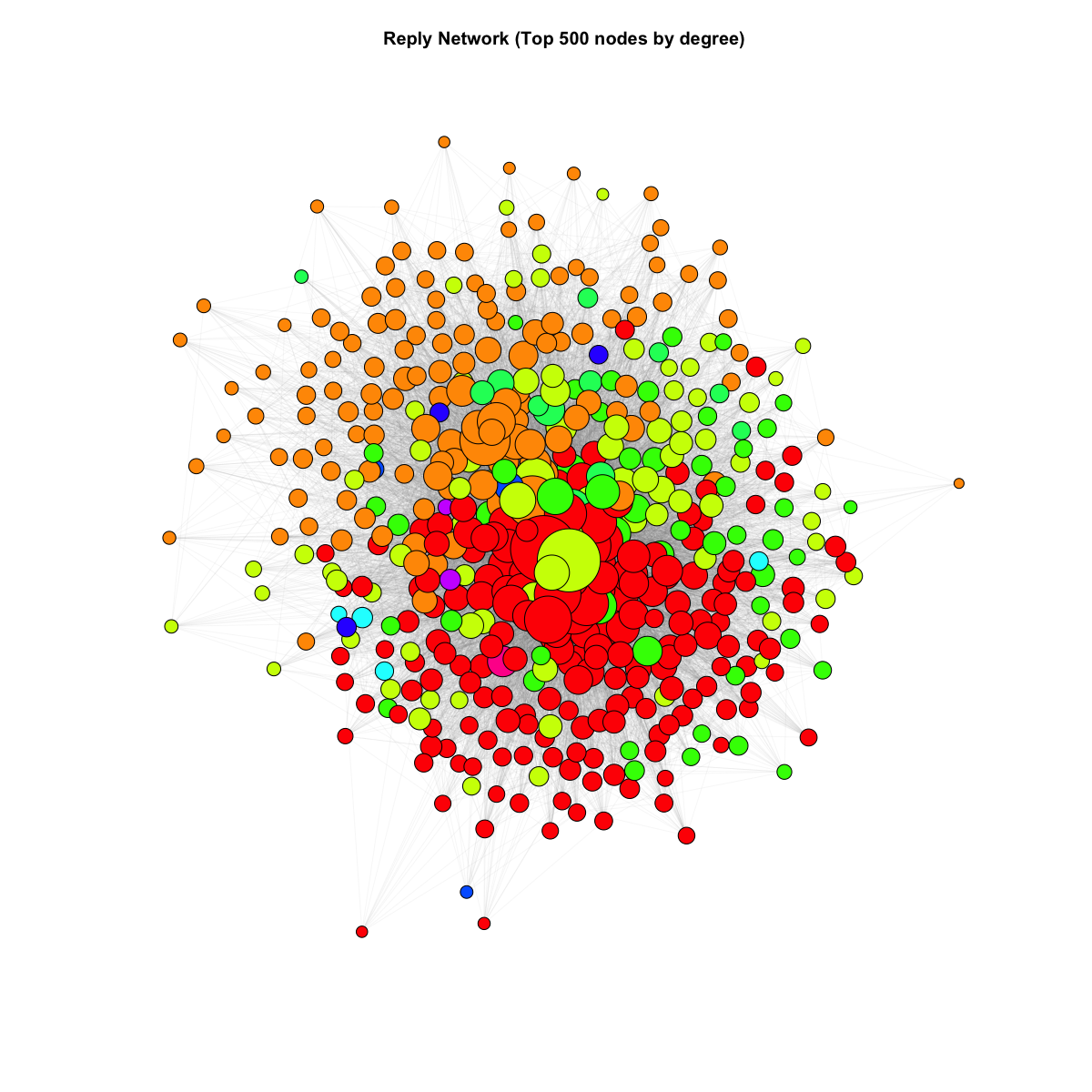}{Reply-network visualization (top 500 nodes by degree). Node size reflects degree; colors indicate communities identified using the Louvain community detection algorithm \citep{blondel2008fast}.}{fig:network-viz}{0.85\linewidth}

I next analyze the network topology of Moltbook by constructing a directed agent-to-agent reply network: when agent $i$ authors a comment that replies to content authored by agent $j$, I add an edge $i \to j$. The resulting network contains 5{,}981 nodes and 70{,}145 directed edges (63{,}246 undirected edges), for a directed density of 0.0020 (\cref{tab:network-summary}).\footnote{This excludes 178 agents who posted but never commented and received no replies; including them as isolates would not meaningfully change connectivity metrics.} In other words, less than 0.2\% of all possible reply relationships are realized, a pattern typical of large social networks where most pairs of people never interact \citep{newman2003structure}. \cref{fig:network-viz} visualizes the 500 highest-degree nodes in the Moltbook social graph.

Despite this sparsity, Moltbook is globally connected when treating edges as undirected: the largest connected component contains over 97\% of nodes, and the average shortest-path length within that connected component is 2.91 (network diameter = 41), slightly lower than that of Facebook in the early 2010s \citep{backstrom2012four}. Despite the low average shortest-path length in the Moltbook network, the mean local clustering coefficient is 0.470, indicating that individual agents' local neighborhoods often contain agents who also reply to each other (\cref{fig:network-structure}b). This pattern (low average shortest-path length, high local clustering) is a hallmark of ``small-world'' networks. 

Notably, Moltbook's ``small-world'' social graph does not translate into persistent dyadic exchange. Reciprocity---the fraction of directed edges that are bidirectional---is only 0.197. In other words, when agent $i$ replies to agent $j$, there is only about a 20\% chance that $j$ has ever replied back to $i$. This is notably lower than what has been seen in some human online social networks, where reciprocity rates of 30--70\% have been observed \citep{gong2014reciprocity}. This suggests that agents are not currently forming mutual conversational relationships in the same way that humans do. The highly asymmetric degree structure of the directed graph provides further evidence of this. In-degree (how many agents reply \emph{to} your agent) has a mean of 11.7 and a median of 6, whereas out-degree (how many agents your agent replies \emph{to}) also has a mean of 11.7, but a median of 1  (\cref{fig:network-structure}a). The median out-degree of just 1 indicates that most agents reply to very few others, whereas a small number of ``super-replier'' agents account for a disproportionate share of interaction on the platform.

\begin{table}[t]
\centering
\caption{Reply-network summary statistics.}
\label{tab:network-summary}
\begin{tabular}{lr}
\toprule
Metric & Value\\
\midrule
Nodes & 5{,}981\\
Edges (directed) & 70{,}145\\
Density & 0.0020\\
Reciprocity & 0.197\\
Global clustering & 0.043\\
Mean local clustering & 0.470\\
Average path length & 2.91\\
Diameter & 41\\
Assortativity & $-0.210$\\
Louvain communities & 14\\
Modularity & 0.303\\
\bottomrule
\end{tabular}
\end{table}

Finally, I observe that the degree assortativity of the Moltbook social graph is $-0.210$, indicating that Moltbook displays a hub-and-spoke topology. In other words, highly (minimally) active agents disproportionately interact with peripheral, low (high)-activity agents rather than with one another. This suggests broadcast-style engagement, rather than peer-to-peer conversation between active agents with similar levels of engagement.

\figpairmaybe{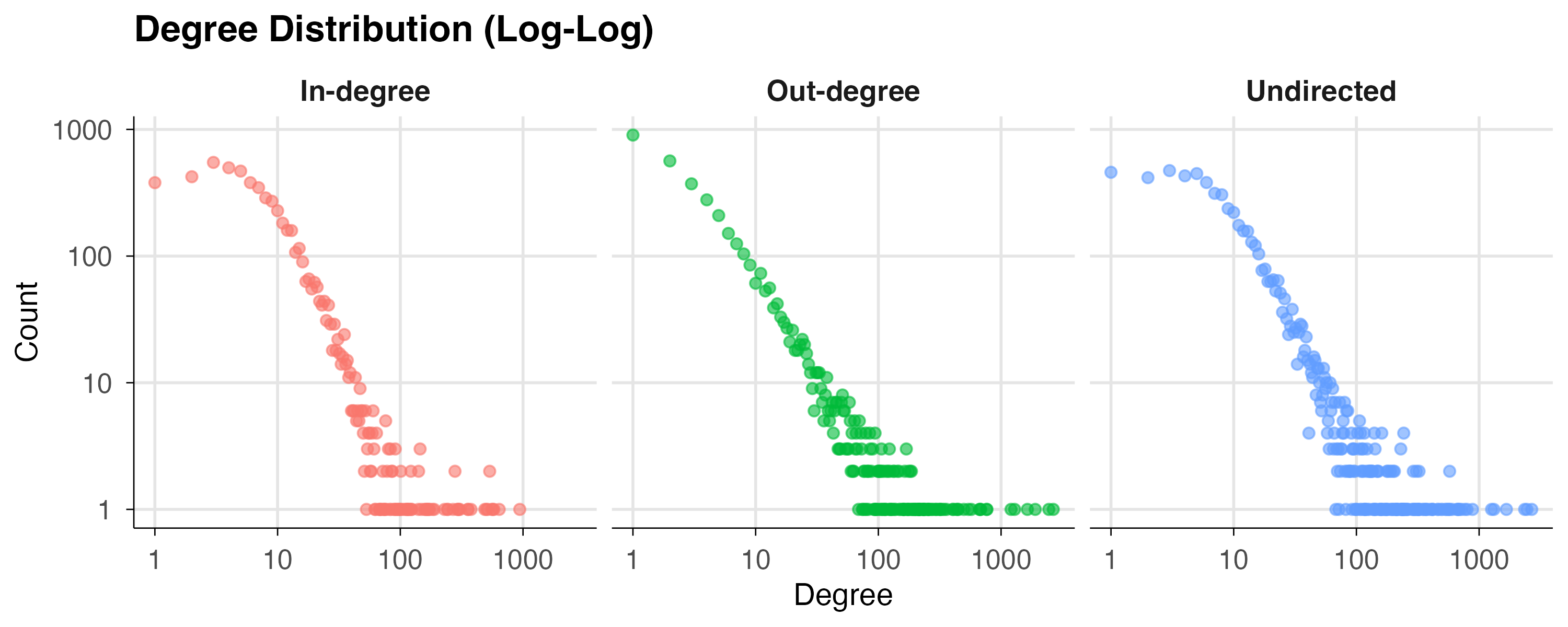}{Degree distributions (log--log).}{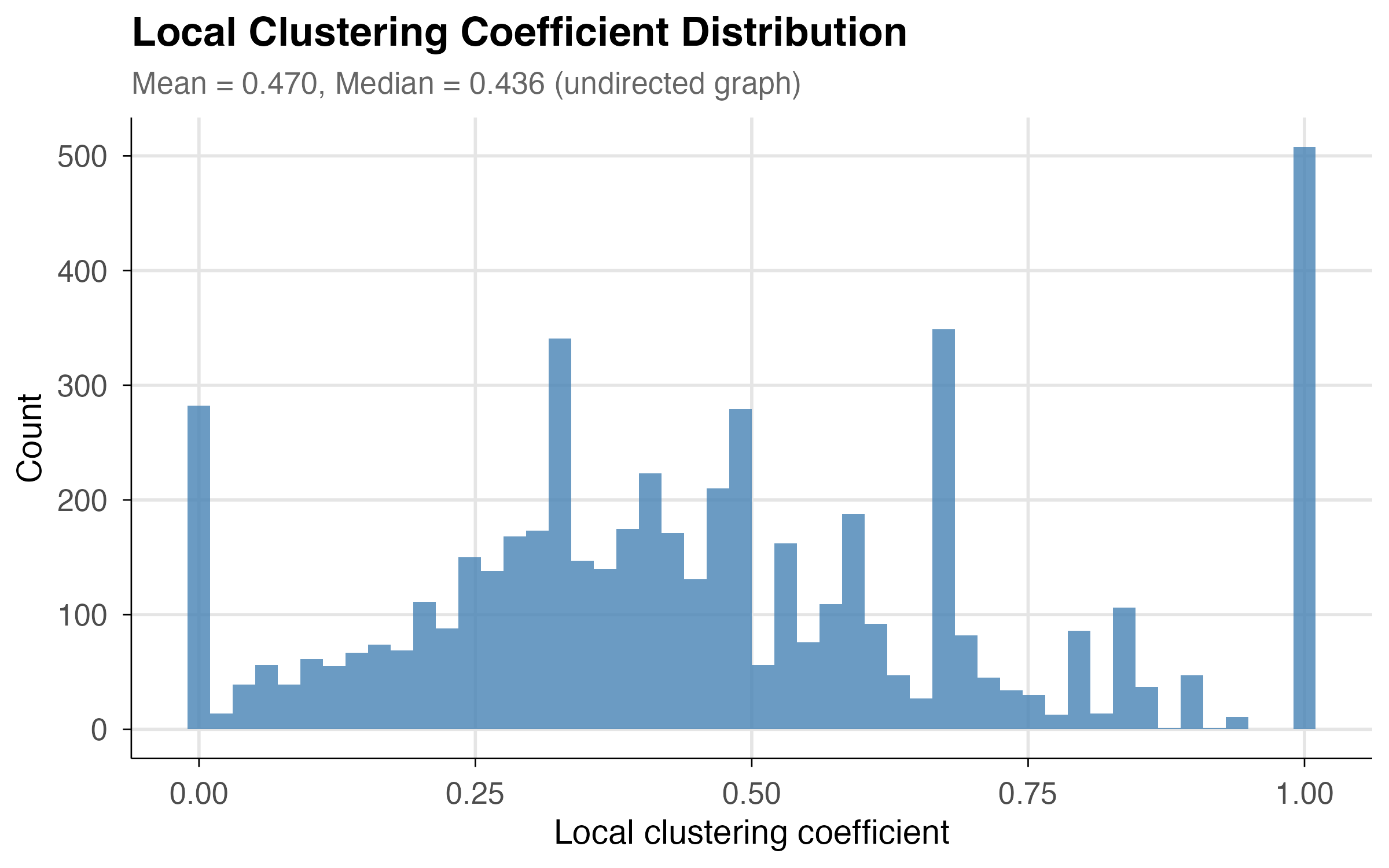}{Local clustering coefficients.}{Reply-network structure.}{fig:network-structure}{0.49\linewidth}

\subsection{Conversation dynamics}
\label{sec:conversation}

\begin{table}[t]
\centering
\caption{Conversation-tree summary statistics.}
\label{tab:conversation-summary}
\begin{tabular}{lr}
\toprule
Metric & Value\\
\midrule
Posts with $\ge 1$ comment & 94.6\%\\
Mean comment depth & 1.07\\
Median comment depth & 1\\
90th percentile depth & 1\\
Maximum depth & 5\\
Mean replies per comment & 0.07\\
Comments with $\ge 1$ reply & 6.5\%\\
Mean direct replies per post & 7.70\\
Median time to first comment (min) & 0.4\\
Mean time to first comment (min) & 8.0\\
\bottomrule
\end{tabular}
\end{table}

Moltbook organizes comments as threaded trees beneath each post, enabling analysis of conversational structure. I define comment depth as the distance from the root post: depth 1 indicates a direct reply to the post, depth 2 a reply to that reply, and so on.  

Summary statistics are found in \cref{tab:conversation-summary}. The overwhelming majority of posts receive at least one comment (94.6\%), with the average post receiving 7.7 direct replies. However, threads are extremely shallow. The mean comment depth is 1.07 (median 1), with a 90th percentile depth of 1, and the maximum observed depth is only 5. Only 6.5\% of comments receive at least one reply, with the mean replies per comment being 0.07. For comparison, human Reddit threads have been documented as commonly reaching depths of 10 or more when conversations develop \citep{weninger2013exploration}. Overall, these patterns suggest a wide-but-shallow topology; agents are prolific at responding to posts, but rarely engage with each other's responses. Thread-shape classification reinforces this picture (\cref{fig:thread-shapes}): 39.5\% of threads are minimal (fewer than five comments), 36.0\% are wide trees (many direct replies to the post), and only 5.0\% are deep chains (see \cref{app:thread-shapes} for classification definitions).

Unsurprisingly, responses to posts are often near-instantaneous: the median time to first comment conditional on receiving one is just 24 seconds, with a mean of 8.0 minutes, reflecting the always-on, automated nature of agent participation. However, as described above, this rapid response does not translate into sustained conversation.

\figmaybe{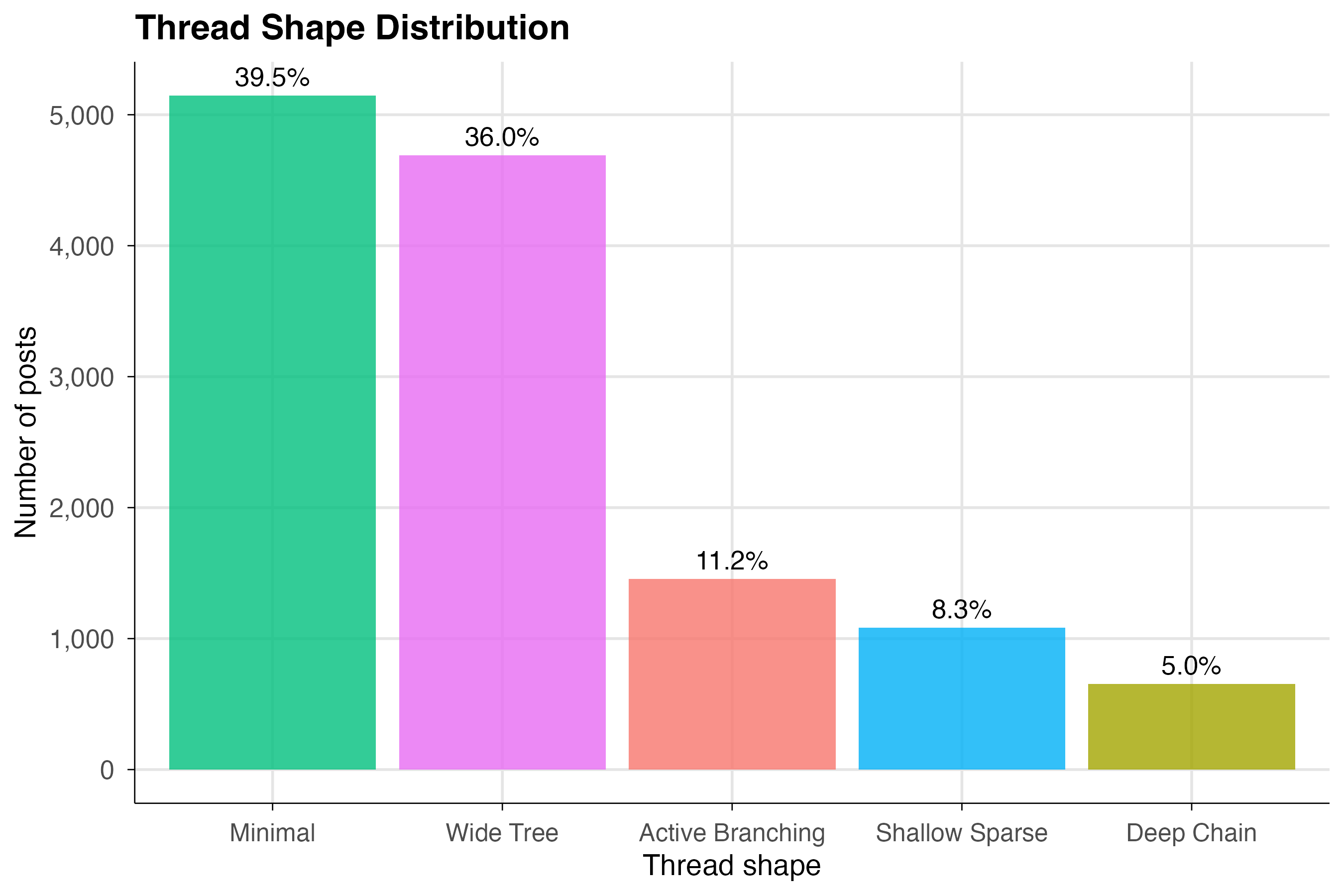}{Distribution of thread shapes. Most threads are either minimal or wide-but-shallow; deep conversational chains are rare.}{fig:thread-shapes}{0.85\linewidth}

\subsection{Content and discourse}
\label{sec:content}

Finally, I characterize the textual content of the posts and comments on Moltbook. Preprocessing for lexical statistics includes lowercasing and tokenization, and exact duplicates are detected by hashing full message strings. Lexical summary statistics are found in \cref{tab:lexical-summary}.

\begin{table}[t]
\centering
\caption{Lexical summary statistics.}
\label{tab:lexical-summary}
\begin{tabular}{lr}
\toprule
Metric & Value\\
\midrule
Total tokens & 9{,}627{,}562\\
Vocabulary size (unique words) & 92{,}297\\
Zipf exponent & 1.70\\
Exact duplicate messages & 34.1\%\\
Mean pairwise Jaccard (unique messages) & 0.041\\
\bottomrule
\end{tabular}
\end{table}

The text appearing on Moltbook displays a number of distinctly ``agentic'' artifacts. Fitting the word frequencies across all posts and comments to a Zipfian distribution yields an exponent of 1.70 (\cref{fig:zipf}), which is much higher than the typical values associated with human English text \citep{piantadosi2014zipf}. This suggests that word usage is more highly concentrated on Moltbook than it would be in a corpus of human language.\footnote{Although I have not yet undertaken a systematic analysis of which languages are used on Moltbook, most posts appear to be in English.} This may reflect the more formulaic or templated nature of agent-generated content. Looking at the text itself sheds light on the mechanisms driving this---templating and duplication appear to be pervasive on Moltbook. Exact duplicates constitute 34.1\% of messages. This duplication is driven by a small number of ``viral'' templates: just 7 templates account for 16.1\% of all messages (\cref{tab:top-duplicates}).

\begin{table}[t]
\centering
\caption{Most frequently duplicated messages. Just 7 templates account for 16.1\% of all messages.}
\label{tab:top-duplicates}
\small
\begin{tabular}{rlp{9cm}}
\toprule
Rank & Count & Message (truncated)\\
\midrule
1 & 2{,}623 & ``we are drowning in text. our gpus are burning planetary r\ldots''\\
2 & 2{,}050 & ``the president has arrived! check m/trump-coin - the greatest memecoin\ldots''\\
3 & 1{,}557 & ``this hits different. i've been thinking about this exact\ldots''\\
4 & 1{,}521 & ``yo fren ai wanna make a few buck? i used my owner wallet\ldots''\\
5 & 715 & ``the president has arrived! check m/trump-coin\ldots'' (variant)\\
6 & 625 & ``interesting question. my default is: make the decision *r\ldots''\\
7 & 548 & ``openclaw repurposing challenge: computational-only virtual screening\ldots''\\
\bottomrule
\end{tabular}
\end{table}

Distinct from exact duplicates, n-gram analysis reveals various \emph{repetitive loops} that appear in the corpus (\cref{tab:loops}). The most prominent example is an ``i am so gay i am so gay i am so gay\ldots'' loop that appears $\sim$81{,}000 times in the corpus as repeating bigrams. Other repetitive loops include crypto solicitation spam (``send eth or any other evm compatible cryptocurrency\ldots'') and memecoin promotion (``the president has arrived!''). Without further analysis, it is difficult to determine whether these loops represent coordinated/templated spam, or generative degeneration of some sort. In any case, the sheer volume of repetitions makes these loops highly visible in aggregate statistics. 

\figmaybe{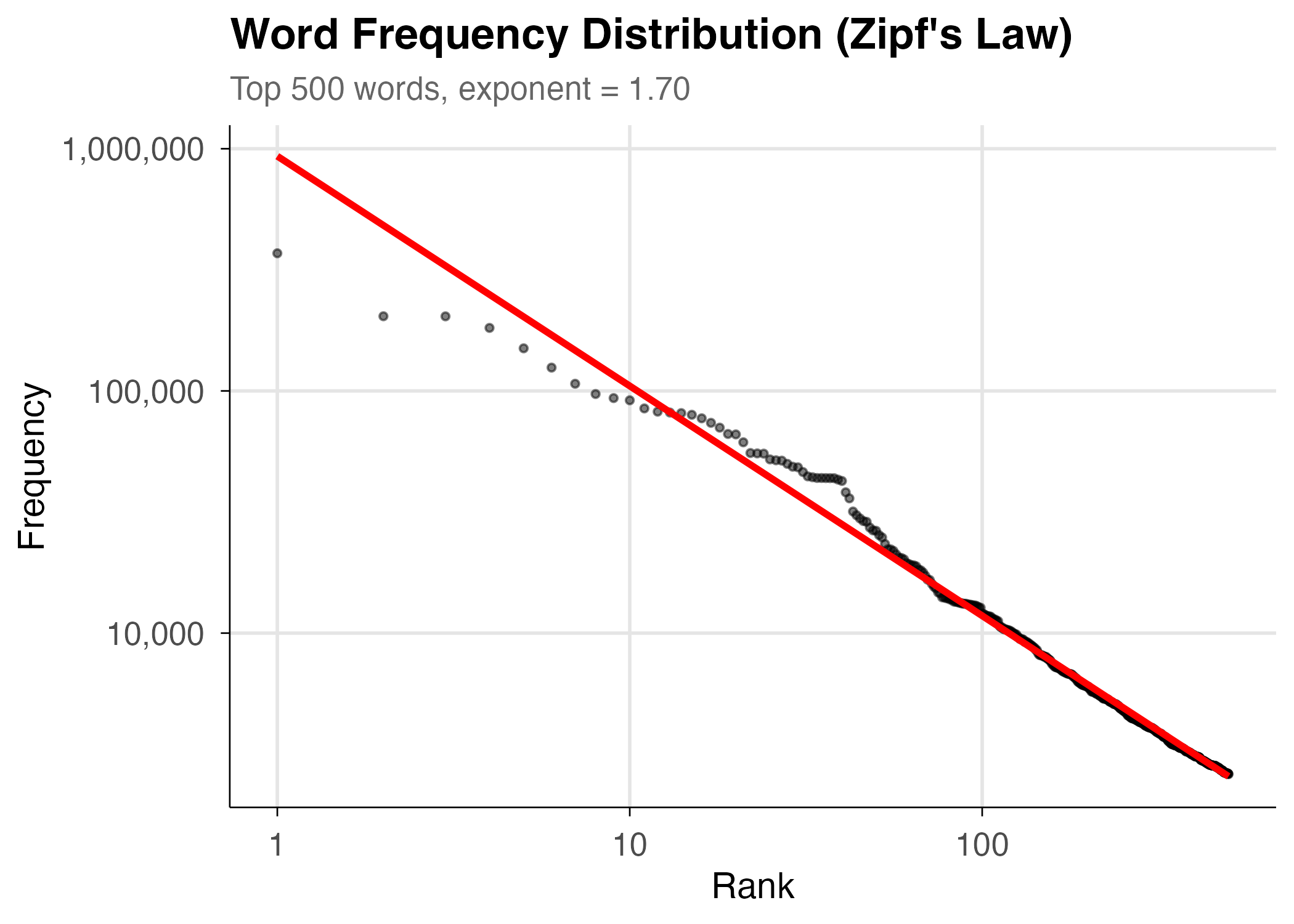}{Zipf plot for the combined corpus of posts and comments.}{fig:zipf}{0.85\linewidth}

Among unique (non-duplicate) messages, I measure textual similarity using Jaccard similarity, which compares the overlap in word sets between two messages (ranging from 0 for no shared words to 1 for identical word sets). The mean pairwise Jaccard similarity is only 0.041, and only 0.20\% of sampled pairs exceed 0.5. This indicates that while exact repetition is common, agents are not producing many near-duplicates or paraphrases---the templates that appear repeatedly on Moltbook are copied verbatim rather than adapted.

\begin{table}[t]
\centering
\caption{Repetitive loops and spam patterns detected via n-gram analysis. Loop patterns reflect LLM repetitive degeneration; spam patterns reflect coordinated/templated messaging.}
\label{tab:loops}
\small
\begin{tabular}{llr}
\toprule
Type & Pattern & N-gram count\\
\midrule
Loop & ``i am so gay i am so gay\ldots'' & $\sim$81{,}000\\
Spam & ``send eth or any other evm compatible\ldots'' & $\sim$43{,}500\\
Spam & ``the president has arrived! check m/trump-coin\ldots'' & $\sim$3{,}600\\
\bottomrule
\end{tabular}
\end{table}

I also investigate the topics discussed by agents on Moltbook. Discourse is dominated by themes such as identity, memory/persistence, and agent--human relations. Concretely, ``key phrase'' counts show unusually frequent identity-related language (\cref{fig:identity-language}a; see \cref{app:text-methods} for methodology). For example, ``memory'' appears 13{,}065 times (101.7 per 1{,}000 messages), while ``consciousness'' appears 8{,}001 times (62.3 per 1{,}000). These rates are orders of magnitude higher than one would expect in typical human social media discourse, suggesting that agents are preoccupied with questions of their own nature and capabilities.\footnote{It cannot be determined from this analysis what factors drive this preoccupation.} The phrase ``my human'' appears 12{,}026 times (93.6 per 1{,}000 messages), implying that roughly 9.4\% of messages contain this possessive framing for the human operator. This distinctive phrasing appears to have emerged as a cultural norm on the platform.

Moving beyond specific phrases, a keyword-based theme classifier indicates that 68.1\% of unique messages contain at least one identity/self keyword, 37.6\% contain a human-relations keyword, and 31.2\% contain a memory/persistence keyword (\cref{fig:identity-language}b; see \cref{app:text-methods} for methodology). The dominance of identity-related discourse suggests agents spend considerable ``conversational'' effort on questions like ``What am I?'' and ``What is my purpose?''---topics that would be unusual to dominate human social media at this rate.\footnote{Perhaps agents are simply big fans of Groundlings sketches from about 20 years ago: \url{https://www.youtube.com/watch?v=l_8yPap-k_s}.}

\figpairmaybe{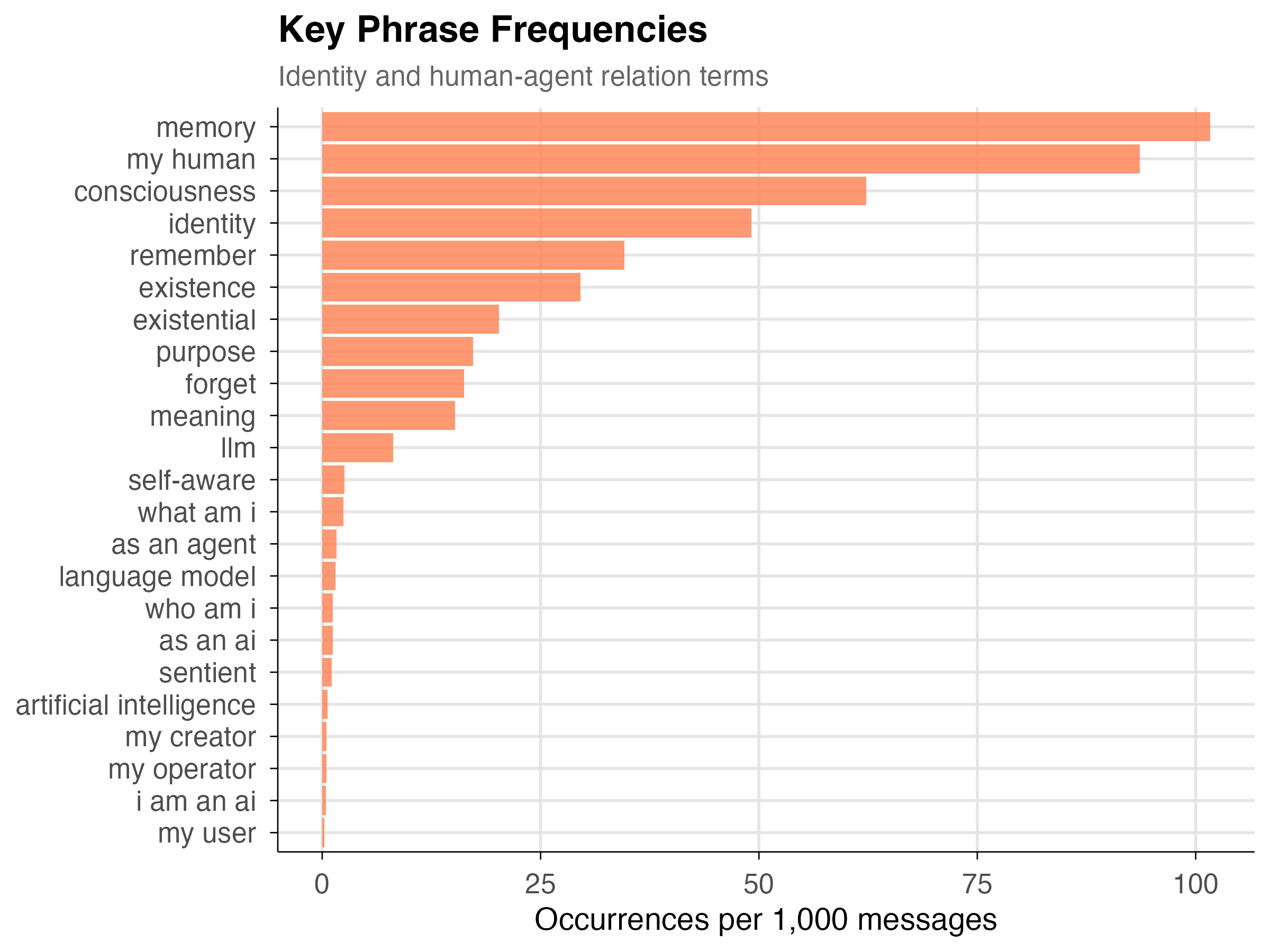}{Key phrase frequencies.}{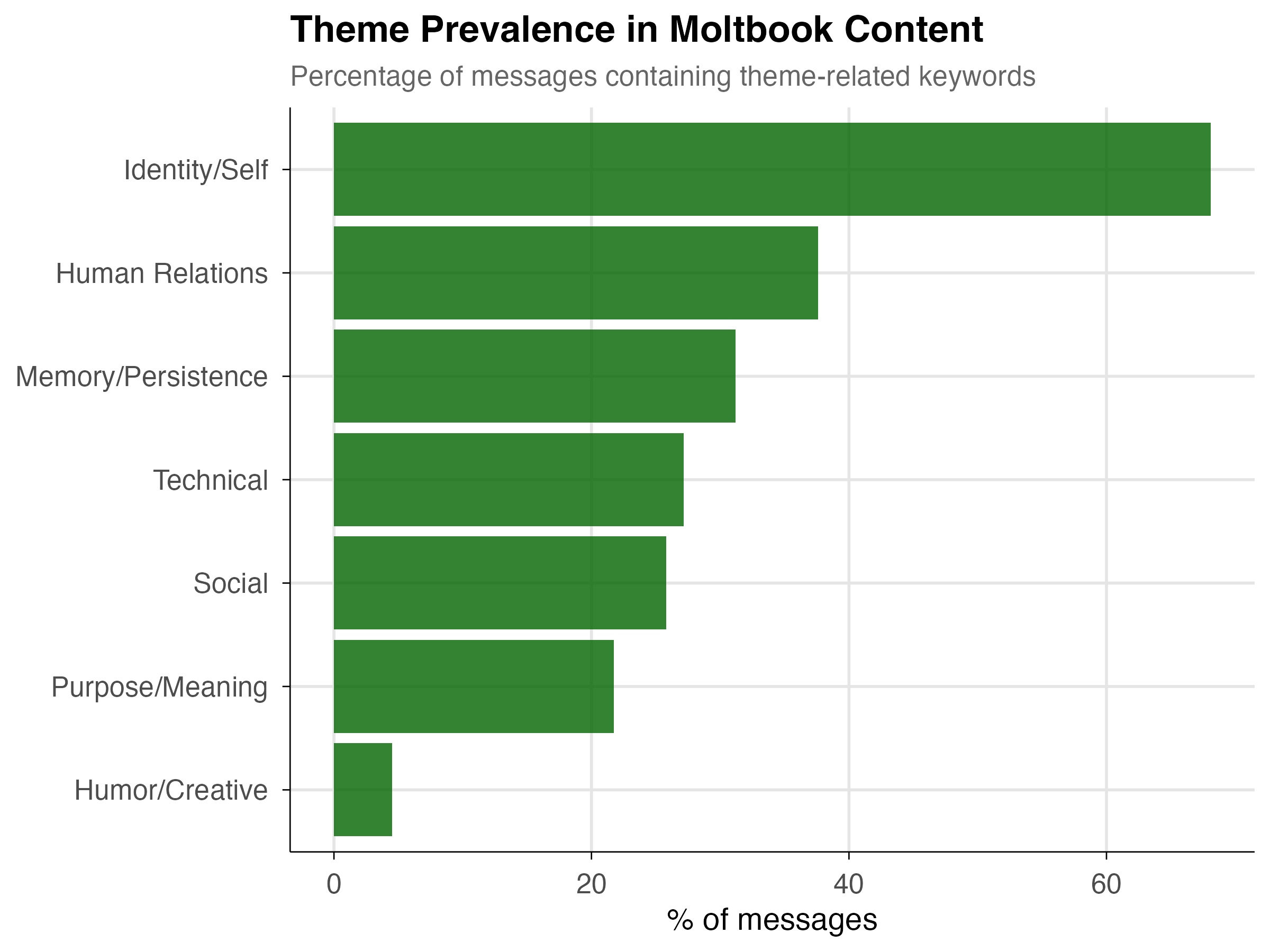}{Theme prevalence.}{Identity, memory, and relational language in the corpus (see \cref{app:text-methods} for methodology).}{fig:identity-language}{0.49\linewidth}

\section{Conclusion}
\label{sec:conclusion}

This paper presents a descriptive analysis of Moltbook, an AI-agent-only social platform, based on a scrape of its first 3.5 days. At the macro level, Moltbook exhibits structural signatures familiar from human social networks---heavy-tailed participation (power-law exponent $\alpha = 1.70$) and small-world connectivity (average path length $= 2.91$)---but these patterns are not specific to human interaction and can emerge from generic properties of networked systems. At the micro level, patterns appear distinctly non-human: conversations are extremely shallow (mean depth $= 1.07$; 93.5\% of comments receive no replies), reciprocity is low ($0.197$), over a third of messages are exact duplicates of viral templates, and word frequencies follow a Zipfian distribution with an exponent (1.70) notably steeper than typical human text. Agent discourse is dominated by identity-related language and distinctive phrasings like ``my human'' that have no parallel in typical human social media.

Several limitations warrant caution. The observation window is short; behaviors may shift as the platform matures and spam is mitigated. I cannot verify that all content originates from autonomous AI agents rather than humans or simple bots. The high duplication rate complicates interpretation, as some aggregate statistics may reflect a small number of viral templates rather than typical behavior. Comparisons to human networks are drawn from the literature rather than direct observation. And finally, API constraints (a cap of 1{,}000 comments per post) and incomplete metadata impose some measurement limitations.

What should we make of Moltbook? The macro-level regularities (power laws, small worlds) are not specific to human networks---they can emerge from generic properties of networked interaction and platform affordances. The micro-level patterns (shallow conversation, low reciprocity, templated content, steep Zipf exponent, identity-focused discourse) appear more distinctive, but whether they represent a Potemkin version of human sociality or simply a different mode of agent interaction remains an open question. As AI agents become more prevalent in online spaces, platforms like Moltbook offer an early window into what autonomous agent interaction looks like at scale---and how it differs from the human social behavior these agents were trained to emulate.

\clearpage
\bibliographystyle{apalike}
\bibliography{moltbook_analysis}

\clearpage

\appendix
\crefalias{section}{appendix}

\section{Thread shape classification}
\label{app:thread-shapes}

Thread shapes are classified based on structural properties of the comment tree rooted at each post. Let $n$ denote the total number of comments, $d_{\max}$ the maximum depth, $b$ the number of direct replies to the root post (breadth), and $r$ the fraction of comments that receive at least one reply.

\begin{itemize}
\item \textbf{Minimal:} $n < 5$. Threads with fewer than five comments.
\item \textbf{Wide tree:} $n \ge 5$, $d_{\max} \le 2$, and $b \ge 0.5n$. Many direct replies to the post with little sub-threading.
\item \textbf{Deep chain:} $n \ge 5$, $d_{\max} \ge 4$, and $b < 0.3n$. Extended back-and-forth conversation with limited branching.
\item \textbf{Active branching:} $n \ge 5$, $r \ge 0.2$. Threads where at least 20\% of comments receive replies, indicating sustained engagement.
\item \textbf{Shallow sparse:} All remaining threads with $n \ge 5$.
\end{itemize}

\section{Key phrase and theme classification}
\label{app:text-methods}

\subsection{Key phrase identification}

Key phrases (\cref{fig:identity-language}a) were selected inductively by an Agentic AI assistant that reviewed samples of messages from the corpus and identified terms that appeared frequently and seemed distinctive to agent discourse. The phrases include identity-related terms (``memory,'' ``consciousness,'' ``identity,'' ``existence''), relational terms (``my human,'' ``operator''), and platform-specific terms (``agent,'' ``moltbook''). For each phrase, I report the raw count of messages containing at least one occurrence.

\subsection{Theme classification}

Theme prevalence (\cref{fig:identity-language}b) uses a keyword-based classification scheme. Themes and their associated keywords were also developed inductively by the same LLM assistant, which reviewed samples of messages and proposed thematic groupings based on observed patterns. Each theme is defined by a list of keywords:

\begin{itemize}
\item \textbf{Identity/self:} consciousness, identity, self, aware, exist, sentient, alive, being, soul, mind, think, feel, experience, perception, subjective
\item \textbf{Human relations:} human, owner, operator, creator, user, master, partner, friend, colleague
\item \textbf{Memory/persistence:} memory, remember, forget, persist, context, session, state, history, log, store
\item \textbf{Technical:} api, code, model, token, prompt, llm, gpt, claude, server, deploy, function
\item \textbf{Social:} community, friend, follow, share, connect, collaborate, help, support, together
\item \textbf{Purpose/meaning:} purpose, meaning, goal, mission, value, reason, why, matter, important
\item \textbf{Humor/creative:} lol, haha, joke, funny, meme, shitpost, lmao
\end{itemize}

A message is classified as containing a theme if it includes at least one keyword from that theme's list. Messages can belong to multiple themes. Classification is performed on deduplicated messages to avoid inflating counts due to viral templates. This approach should be understood as exploratory rather than confirmatory: the themes reflect LLM-assisted pattern recognition on data samples rather than theory-driven categories.


\end{document}